\documentclass[useAMS,usenatbib]{mnras}

\usepackage{color}
\usepackage[english]{babel}
\usepackage{graphicx}
\usepackage{amsmath}
\usepackage{amssymb}
\usepackage{epsf}
\usepackage{bm}
\usepackage[utf8]{inputenc}
\begin{document}

\def\pFn{p_{\rm Fn}}
\def\pFp{p_{\rm Fp}}
\def\pFe{p_{\rm Fe}}

\newcommand{\om}{\mbox{$\omega$}}              
\newcommand{\Om}{\mbox{$\Omega$}}              
\newcommand{\Th}{\mbox{$\Theta$}}              
\newcommand{\ph}{\mbox{$\varphi$}}             
\newcommand{\del}{\mbox{$\delta$}}             
\newcommand{\Del}{\mbox{$\Delta$}}             
\newcommand{\lam}{\mbox{$\lambda$}}            
\newcommand{\Lam}{\mbox{$\Lambda$}}            
\newcommand{\ep}{\mbox{$\varepsilon$}}         
\newcommand{\ka}{\mbox{$\kappa$}}              
\newcommand{\dd}{\mbox{d}}                     
\newcommand{\nb}{\mbox{$n_{\rm b}$}}                     

\newcommand{\Tb}{\mbox{$T_{\rm b}$}}         
\newcommand{\kB}{\mbox{$k_{\rm B}$}}           
\newcommand{\tn}{\mbox{$T_{{\rm c}n}$}}        
\newcommand{\tp}{\mbox{$T_{{\rm c}p}$}}        
\newcommand{\te}{\mbox{$T_{eff}$}}             
\newcommand{\ex}{\mbox{\rm e}}                 
\newcommand{\rate}{\mbox{${\rm erg~cm^{-3}~s^{-1}}$}}
\newcommand{\gcc}{\mbox{${\rm g~cm^{-3}}$}}
\newcommand{\xr}{\mbox{$x_{\rm r}$}}
\newcommand{\gs}{\mbox{$g_{\rm s}$}}
\newcommand{\me}{\mbox{$m_{\rm e}$}}
\newcommand{\msun}{\mbox{${\rm M}\odot$}}
\def\mg{\textcolor{magenta}}



\title[Deformed crystals and oscillations of neutron stars]{Deformed crystals and 
torsional oscillations of neutron star crust}

\author[A. A. Kozhberov, D. G. Yakovlev ]{{A. A.    Kozhberov\thanks{E-mail: kozhberov@gmail.com}},
     {  D. G. Yakovlev}\\
    Ioffe Institute, Politekhnicheskaya 26, St~Petersburg 194021, Russia\\
    }

\maketitle \label{firstpage}

\begin{abstract} 
We study breaking stress of deformed Coulomb crystals in a neutron star crust, taking
into account electron plasma screening of ion-ion interaction;
calculated breaking stress is fitted as a function of electron screening
parameter. We apply the results for analyzing
torsional oscillation modes in the crust of a non-magnetic star. 
We present exact analytic expression for the fundamental frequencies
of such oscillations and show 
that the frequencies of all torsional oscillations are insensitive to
the presence of the outer neutron star crust. The results can be useful in theoretical 
modeling of processes involving deformed Coulomb crystals
in the crust of neutron stars, such as magnetic field
evolution, torsional crustal or magneto-elastic quasi-periodic
oscillations of flaring soft gamma-ray repeaters, pulsar glitches. 
The applicability of the results to soft gamma-ray repeaters is discussed.    
\end{abstract}



\section{Introduction}
\label{s:introduc}

Neutron stars are thought to have a bulky and
massive liquid core of superdense matter \citep[e.g.,][]{ST1983}.
The core is surrounded by a thin and light envelope (with thickness 
$\sim 1$ km and with mass of $\sim 0.01\,\msun$) which is often called the
crust; its matter contains
ions (atomic nuclei), electrons and possibly free neutrons. The ions constitute the Coulomb
plasma (e.g. \citealt{HPY2007}) which is typically in crystalline state but
may be liquid or gaseous especially in the surface
layers of warm stars.

The crust is known
to be divided into the outer crust ($\rho \leq \rho_{\rm drip} = 4.3 \times 
10^{11}$ \gcc) and the deeper inner crust. The outer crust consists predominantly
of the ions and electrons. At $\rho \gtrsim 10^6$ \gcc\ (about 10 m under
the neutron star surface) the ions become fully
ionized and the electrons become strongly degenerate and relativistic. 
In the inner crust, free neutrons
appear in addition to the ions and electrons; they are dripped off the nuclei
and further complicate nuclear physics of the crust. The free neutrons are typically
degenerate and superfluid.

Crystalline ions in the  crust can be described by the Coulomb crystal model. 
This model is also used in the physics of white dwarfs as well as in the physics of dusty
plasmas, available in space and laboratory, with numerous applications in science and technology
(see, e.g., \citealt{2010Vaulina} and references therein). 
Here we primarily consider neutron star matter. 
To study Coulomb lattice, one commonly 
employs a model of ion-ion interaction screened by polarization of plasma electrons
(as detailed in Section \ref{s:mainpar}). 
The electron screening is relatively weak through the crust bulk, at
$\rho \gtrsim 10^{6}$ \gcc, although one usually takes it into account 
(e.g., \citealt{CH10}). At smaller densities, the electron plasma screening
can be stronger but the electron screening model can be more complicated
because of partial ionization of ions.

When neutron stars evolve, their crust undergoes various deformations
\citep[e.g.,][]{O05,CH08,BK17,FHL18,2018Gabler,BC18}. For magnetars, these deformations
are primarily associated with the magnetic field \citep[e.g.,][]{BL14,L16,LL16};
for pulsars they are thought to be connected with glitches \citep[e.g.,][]{PFH14}.
Investigations of deformed lattices are important for understanding different
processes in neutron star interiors. 

In Section \ref{s:deformed} we consider properties of deformed Coulomb crystals with the
focus on the effect of electron plasma screening of the ion-ion interaction
on the breaking stress. Stronger screening simplifies the breaking; one 
should study the strength of this effect for modeling various phenomena associated with deformations
of Coulomb crystals. We calculate the breaking stress of mono-crystals under some 
specific deformations and approximate the results by an accurate analytic function
of the electron screening parameter. We outline further breaking stress of 
poly-crystals (isotropic solids); this model is common in neutron star physics.

In Sections \ref{s:torsion} and \ref{s:properties} we investigate torsional oscillations 
of neutron stars which
are determined by elastic properties of Coulomb crystals. Torsional oscillations 
are widely used to explain quasi-periodic oscillations (QPOs) in the
spectra of soft gamma-ray repeaters (SGRs); see e.g. 
\citet{2006Watts,2014Huppenkothen,2014Huppen}. There are many theoretical 
investigations of torsional (crustal elastic) and mageto-elastic oscillations of neutron stars
aimed at explaining observed QPO frequencies; e.g.,  
\citet{2016Gabler,2018Gabler}
and references therein. 

We will study torsional oscillations of non-rotating and non-magnetic stars. In spite
of many publications, not all their properties have been investigated in detail. Our
primary interest will be in analytic analysis of oscillation spectrum,  
energy, breaking conditions and special behavior of outer crystallized
layers in these oscillations. We discuss and conclude in Section~\ref{s:conclude}.

\section{Deformed Coulomb crystals in neutron stars}
\label{s:deformed}

\subsection{Main parameters}
\label{s:mainpar}

In this section we summarize the data on deformed Coulomb crystals of
atomic nuclei in a neutron star crust paying special attention on the
maximum (breaking) stress. We will consider widely used models of 
the crust which contains ions (spherical atomic nuclei) of one type
in a given matter element. Type and properties of atomic nuclei
change with the density $\rho$ (e.g., \citealt{HPY2007}), 
from ordinary nuclei (such as
$^4$He, $^6$C, $^{26}$Fe) near the surface to very neutron rich nuclei
at the crust-core interface (at $\rho_{\rm cc} \approx 1.5 \times 10^{14}$ 
\gcc) which would be highly unstable in laboratory.  Variation 
of atomic nucleus parameters with depth is a complicated phenomenon governed by
nuclear physics and by pre-history of crustal matter (accretion,
cooling, nuclear burning, beta-processes). 

In spite of the complexity of nuclear physics, the description 
of Coulomb crystals of atomic nuclei is based on a few simple
parameters (e.g., \citealt{HPY2007}),
\begin{equation}
   \Gamma=\frac{Z^2 e^2}{a_{\rm i}\kB T}, \quad
   a_{\rm i}=\left(\frac{3}{4 \pi n_{\rm i}} \right)^{1/3},
\label{e:Gamma}   	
\end{equation}	 
where $\Gamma$ is the Coulomb coupling
parameter, $n_{\rm i}=N_{\rm i}/V$ is the number density of the ions, $Z$
is the charge number of one ion, $N_{\rm i}$ is the number of ions
(in a volume $V$), 
$a_{\rm i}$ is the ion-sphere (Wigner-Seitz) radius,
$T$ the temperature, $e$ the elementary charge, and 
$\kB$ the Boltzmann constant. A classical system of 
the ions solidifies at $\Gamma \geq \Gamma_{\rm m}\approx 175$,
i.e. at $T \leq T_{\rm m} \approx Z^2 e^2/(\kB a_{\rm i}\Gamma_{\rm m})$.

While simulating  Coulomb plasmas of ions
one often uses the model of exponentially screened Coulomb
interaction for a pair of ions,
\begin{equation}
U(r)= \frac{Z^2 e^2}{r} \exp(-\kappa r)~,
\label{Yuk}
\end{equation}
where $r$ is a distance between the ions, and $\kappa$ is an
inverse electron plasma screening length. In the very outer 
neutron star layers, where
the electrons are non-degenerate, $\kappa$ is equal to the 
inverse electron Debye screening length. In the deeper layers 
of strongly degenerate electrons, it is the
Thomas-Fermi electron wavenumber,
\begin{equation}
\kappa_{\rm TF}\equiv \sqrt{4\pi e^2 \frac{\partial n_{\rm e}}{\partial\mu_{\rm e}}}
= \frac{p_{\rm F}}{\hbar}\,\left(\frac{4e^2}{{\rm \pi} v_{\rm F}}   \right)^{1/2}, 
\label{e:TF}
\end{equation}
where $n_{\rm e}=Z n_{\rm i}$ is the number density of electrons,
$\mu_{\rm e}$ is the electron chemical potential,
$p_{\rm F}$ the electron Fermi momentum, and $v_{\rm F}$ 
is the electron Fermi velocity. 

We will be mainly interested in these deeper layers. The electron screening
of the inter-ion potential (\ref{Yuk}) can be characterized by
the dimensionless screening parameter $s\equiv\kappa a_{\rm i}$
that is typically small but generally non-negligible.
The case of $s=0$ corresponds to the absence of plasma screening.
For the screening by degenerate electrons, from
equation (\ref{e:TF}) we have
\begin{equation}
   s=\kappa_{\rm TF}a_{\rm i}=0.1850\,Z^{1/3}\,\frac{c}{v_{\rm F}},
\label{e:sTF}   
\end{equation}
where $c$ is the velocity of light. Through the bulk
of the crust, for any composition of the crustal matter (e.g., \citealt{HPY2007})
one has $s \lesssim 0.7$. Higher $s$ may be realized
in a narrow layer near the surface, at $\rho \lesssim 10^6$ \gcc.
However, the applicability of the screened Coulomb potential given by equation (\ref{Yuk}) 
in this layer requires further study because of the effects of partial ionization.

\subsection{Breaking stress of Coulomb crystals}
\label{s:break}

Let us outline theoretical data on the breaking stress 
$\sigma_{\rm max}$ of Coulomb crystals, paying particular attention on its dependence on $s$.

Most of the previous studies of stressed Coulomb crystals have focused on
the shear modulus $\mu$ of these crystals
 \citep[e.g.,][]{OI90,S91,HH08,HK09,H12,CH12}, paying less attention to
their stability. 
Analytically, it has been done by \cite{BK17,BC18} for the bcc lattice
with the uniform electron background. Deformations of lattices with $s > 0$
have been investigated by \cite{CH10,HH12} via molecular dynamic (MD) simulations for
a restricted number of $s$ values; these results will be discussed further.

For obtaining $\sigma_{\rm max}$ we need the potential energy of a Coulomb crustal. 
Let a distance between two ions $i$ and $j$ be $r_{ij}=|\bm{R}_i-\bm{R}_j+\bm{u}_i-\bm{u}_j|$,
where $\bm{R}_i$ is the equilibrium position of the $i$th ion,
and $\bm{u}_i$ is its displacement.
At $\Gamma \gg \Gamma_{\rm m}$, motions of crystalline ions 
can be considered as small oscillations around their equilibrium positions.
Then their potential energy can be expanded in powers of $\bm{u}_i$,
\begin{equation}
U_{\textmd{i}}\approx{U_{\textrm{M}}}+\frac{1}{2}\sum\limits_{i,j=1}^{N_{\rm i}}\
{u_{i}^{\alpha}}{u_{j}^{\beta}}\frac{{{\partial }^{2}}{{U}_{\textmd{i}}}}
{\partial u_{i}^{\alpha}\partial u_{j}^{\beta}}\
{\bigg\vert_{u_{i}^{\alpha},u_{j}^{\beta}=0}}~,\
\label{U02}
\end{equation}
where Greek indices enumerate Cartesian vector components,
and summation over repeated indices is assumed; 
$U_{\rm M}$ is the electrostatic (Madelung) energy of the lattice.

For any lattice with one ion in the elementary cell 
at $s \geq 0$, the electrostatic 
energy can be written as (e.g., \citealt{B02})
\begin{eqnarray}
U_\textrm{M}&=&N_{\rm i}\frac{Z^2e^{2}}{a_{\rm i}}\zeta,\\
\frac{\zeta}{a_{\rm i}}&=&\sum_{l\neq0} \frac{E_{-}+E_{+}}{4R_l}
-\frac{\kappa}{2}{\rm erf}\left(\frac{\kappa^2}{2A^2}\right)
-\frac{A}{\sqrt{\pi}}{\rm e}^{-\frac{\kappa^2}{4A^2}} \nonumber \\
&+&\sum\limits_{m} \frac{2\pi n_{\rm i}}{G_m^2+\kappa^2}
{\rm e}^{-\frac{G_m^2+\kappa^2}{4A^2}}-\frac{2\pi n_{\rm i}}{\kappa^2}~,
\label{UTF}
\end{eqnarray}
where $E_{\pm}={\rm e}^{\pm\kappa R_l}\,{\rm erfc}
\left(AR_l\pm {\kappa}/(2A)\right)$,
$\bm{R}_l$ is a direct lattice vector, 
$\bm{G}_m$ is a reciprocal lattice vector,
${\rm erf}(x)$ is the error function, 
${\rm erfc}(x)\equiv 1-{\rm erf}(x)$,
and $A$ is an arbitrary constant;
$A\approx 2/a_{\rm i}$ is most suitable because
it gives good numerical convergence of the sums.

We assume that the ions form a body-centered cubic (bcc) lattice 
because at $s = 0$. This lattice possesses
the lowest electrostatic energy; its Madelung constant  $\zeta=-0.895929255682$ \citep[e.g.,][]{CF16}.
The dependence of  $\zeta$ on
$s$ is naturally the same for degenerate and non-degenerate
electron background as long as the ion-ion interaction is 
described by equation 
(\ref{Yuk}).
For degenerate relativistic electrons in the envelopes 
of neutron stars, according
to equation (\ref{e:sTF}), we have 
$s \lesssim 0.7$. For non-degenerate (or degenerate but non-relativistic) 
electrons this
criterion can be formally less strict \citep{HF94}, but  
it is limited by partial
ionization of ions and associated screening of the inter-ion
interaction by bound electrons.

Here, we present the first semi-analytical study of
stability of a deformed bcc Coulomb crystal with $s > 0$.
Let us introduce a Cartesian coordinate system with the main lattice cube edges
of the bcc lattice oriented in such a way that the direction to the nearest neighbor
is given by vector $0.5a_{\rm l}(1,1,1)$, $a_{\rm l}$ being the lattice constant.
Following \citet{CH10}, we consider such deformation of this lattice, which translates the
vector $a_{\rm l}(n_1,n_2,n_3)$ as
\begin{equation}
a_{\rm l}(n_1,n_2,n_3)\to
a_{\rm l}\left(n_1+\frac{\epsilon}{2}n_2,
n_2+\frac{\epsilon}{2}n_1,\frac{n_3}{1-\epsilon^2/4}\right),
\label{e:trans1}
\end{equation}
where $\epsilon$ is a small deformation parameter, while $n_1$, $n_2$, and $n_3$ are 
arbitrary integers. 

With increasing $\epsilon$ the crystal becomes more strained and stressed.
It breaks at some critical $\epsilon=\epsilon_\text{max}$ which
depends on $s$. The effective stress at any $\epsilon$ is calculated as
\begin{equation}
   \sigma(s,\epsilon)=\frac{\partial {\cal E}}{\partial \epsilon}
	= n_{\rm i}\frac{Z^2 e^2}{a_{\rm i}}\,\frac{\partial \zeta}{\partial \epsilon},
\label{sig}
\end{equation} 
where we take 
the Madelung energy $U_{\rm M}$ as an internal energy at zero temperature 
and neglect the energy of ion vibrations, so that 
the internal energy density of the ions is ${\cal E}=U_\text{M}/V$.
The breaking stress is then given by  $\sigma_\text{max}(s)\equiv\sigma(s,\epsilon_{\rm max})$
(e.g., \citealt{CH10}). For a crystal with the uniform electron background ($s=0$), 
one has $\epsilon_{\rm max}=0.1109$.

MD studies determine 
$\epsilon_\text{max}$ and $\sigma_{\rm max}$ 
through direct simulations of crystal evolution under
increasing $\epsilon$.

In semi-analytical studies, 
the lattice at low temperatures can be
treated as unstable if one or more of the squared 
frequencies of phonon modes
become negative at some phonon wave vector. 
In other words, in a stable crystal
the second-order term in equation (\ref{U02}) 
should be a positive definite quadratic form.
Increasing the plasma screening $s$ makes the crystals less stable.
For the bcc lattice, there
exist a critical value $s_{\rm max }=4.76$ which 
completely destroys  crystalline state. It was
calculated by \cite{RKG88} in 
MD simulations of the dusty plasma crystals; it  
was independently proven by \cite{Kd18}. It is
not clear if this effect can be realized in the neutron
stars because the applicability of equation (\ref{Yuk})
at so high $s$ in neutron star matter is questionable,
but the effect may take place in dusty plasmas
(see Section 1).

For a given $s$, we have determined $\epsilon_{\rm max}$ as the critical
strain, at which some phonon frequencies become complex numbers, and
obtained then the breaking stress $\sigma_{\rm max}$ from equation (\ref{sig})
at this $\epsilon_{\rm max}$. In Table \ref{Tab} we list $\epsilon_{\rm max}$
and $\sigma_{\rm max}$ for a number of $s$ values (with $s\leq 1.81$). Naturally,
increasing the electron screening reduces $\epsilon_{\rm max}$ and $\sigma_{\rm max}$.

Note that the values of $\sigma_{\rm max}$ from Table \ref{Tab} can be 
accurately approximated as
\begin{equation}
   \sigma_\text{max}(s)=\sigma_\text{max}(0)\,
   \frac{\sqrt{1+1.451\,s^2}}{1+0.755\,s^2}.
\label{e:fit}   
\end{equation}
The root-mean squared relative error of
this fit is 0.7 per cent, and the
maximum error of
1.5
per cent occurs at $s=1.81$. 

Let us stress that we have calculated the 
phonon spectrum in the linear harmonic-lattice approximation.
Such phonon frequencies are independent of temperature, and
the approach is well justified for temperatures $T$, which 
are sufficiently lower than the melting temperature $T_{\rm m}$.
If $T$ is close to $T_{\rm m}$, the linear theory may become 
inaccurate because of anharmonic effects. Therefore, we warn
the reader that our results (Table \ref{Tab}) can be inaccurate near
the melting point.

\renewcommand{\arraystretch}{1.2}
\begin{table}
\caption{Breaking strain $\epsilon_{\max}$ and stress $\sigma_{\rm max}$ versus $s$ at $T=0$ for
transformation (\ref{e:trans1}) of the bcc lattice}
\label{Tab}
	\begin{tabular}{@{}lcclcc@{}}
		\hline
		~~$s$ & $\epsilon_{\max}$ & $\widetilde{\sigma}_{\max}^{a)}$ 
		&	$~~~s$ & $\epsilon_{\max}$ & $\widetilde{\sigma}_{\max}$\\
		\hline
		0.0   & 0.1109  & 0.02007 &		1.3   & 0.0998  & 0.01651 \\
		0.4   & 0.1099  & 0.01972 &		1.42  & 0.0979  & 0.01591 \\
		0.5714 & 0.1090 & 0.01938 &		1.53  & 0.0957  & 0.01529 \\
		0.71  & 0.1079  & 0.01901 &		1.62  & 0.0938  & 0.01477 \\
		0.87  & 0.1059  & 0.01841 &		1.73  & 0.0918  & 0.01417 \\
		1.04  & 0.1038  & 0.01773 &		1.81  & 0.0898  & 0.01366 \\
		1.19  & 0.1018  & 0.01708 &		  &   &   \\
		\hline 
\end{tabular}
\\
$^{a)}~~ \widetilde{\sigma}_{\rm max}
=\sigma_{\rm max}/(n_{\rm i} Z^2e^2/a_{\rm i})$
\end{table}

Now we can compare our results 
with MD simulations 
performed by \citet{CH10}
at the one value of $s=4/7 \approx 0.5714$.
They approximated their results by
\begin{equation}
\sigma_{\max}^{\rm MD}(4/7)=n_{\rm i}\frac{Z^2 e^2}{a_{\rm i}} 
\left(0.0195-\frac{1.27}{\Gamma-71}\right)~.
\label{chu}
\end{equation}
Then at small $T$ their   $\sigma_{\max}^{\rm MD}
=0.0195n_{\rm i}Z^2e^2/a_{\rm i}$ deviates from our
value (line 3 in our Table \ref{Tab}) only by a fraction of per cent.
This good agreement supports practical equivalence of two criteria 
of stability --- the MD one and the one based on complex-valued phonon modes.
Note that \cite{CH10} give the typical breaking strain $\epsilon_{\rm max} \approx 0.13$, 
which approximately agrees with our calculations. A slight difference 
can be explained by the fact that the part of the first Brillouin zone, 
where the complex-valued frequencies appear, 
is small and noticeably increases with $\epsilon$ (see \citealt{BK17}).

We have also considered two other shear deformations of bcc lattice. The
first one is
\begin{equation}
a_{\rm l}(n_1,n_2,n_3)\to a_{\rm l} \left(n_1+n_3{\epsilon},n_2,n_3\right),
\label{e:trans2}
\end{equation}
where $\epsilon$ is again the deformation parameter. We have calculated the breaking
stress $\sigma_{\rm max}$ in the same way as in Table~\ref{Tab}. The results
are presented in Table \ref{Tab2} for $s$=0, 0.5 and 1. They are
accurately fitted by the same equation (\ref{e:fit}), 
which fits the data of Table \ref{Tab}. 
Note that equation (\ref{e:fit}) is not supposed
to be highly accurate to for all shear deformations.

\begin{table}
	\caption{ The same as in Table \ref{Tab} but for the  lattice 
	transformation (\ref{e:trans2})}
	\label{Tab2}
	\begin{tabular}{@{}ccc@{}}
		\hline
		$s$ & ${\epsilon}_{\max}$ & $\widetilde{\sigma}_{\max}$ \\
		\hline
		0 & 0.1051 & 0.01928 \\
		0.5 & 0.1036 & 0.01874 \\
		1 & 0.0990 & 0.01719 \\
		\hline
	\end{tabular}
\end{table}

The next shear deformation is a compression ($\epsilon<0$) 
neglecting the electron screening ($s=0$), at which
\begin{equation}
a_{\rm l}(n_1,n_2,n_3)\to a_{\rm l} 
\left((1+\epsilon) n_1, \frac{n_2}{\sqrt{1+\epsilon}},
\frac{n_3}{\sqrt{1+\epsilon}}\right).
\label{e:trans3}
\end{equation}
According to \cite{BC18} at $\epsilon <0$ it gives one of the lowest strains,
$|\epsilon_{\rm max}|=0.0435$. For this strain we have calculated 
$\partial \zeta / \partial \epsilon=0.00542$, which yields
$\sigma_{\rm max}=0.00542\, n_{\rm i}Z^2 e^2/a_{\rm i}$ from equation (\ref{sig}).
This demonstrates once again the well
known fact that $\sigma_{\rm max}$ and $\mu$ in a Coulomb crystal noticeably  
depend on deformation type (e.g. \citealt{BK17,Kd18,BC18} and references therein).
Note that the reported value of $\sigma_{\rm max}$ is close to the
value of $\mu \,\epsilon_{\rm max} \approx 0.0052\,  n_{\rm i}Z^2 e^2/a_{\rm i}$, where
$\mu$ is given by equation (\ref{e:mu}) presented below.

For practical applications in a neutron star crust we will consider the
model of isotropic solid in which \citep{OI90} 
\begin{equation}
    \mu = 0.1194n_{\rm i}\frac{Z^2 e^2}{a_{\rm i}},
\label{e:mu}    
\end{equation}
and take the maximum strength in the form
\begin{equation}
\sigma_{\max} \approx 0.02n_{\rm i}\frac{Z^2 e^2}{a_{\rm i}}.
\label{e:sigmamax}
\end{equation}
We have not found any deformation in a bcc mono-crystal, 
which gives significantly larger $\sigma_{\rm max}$.
This value agrees with results of \cite{HH12}; in all their simulations 
$\sigma_{\rm max}$ has not been larger.

Please bear in mind that we have mainly discussed simulations
in mono-crystals, whereas formation of polycrystals 
(particularly isotropic solids) in neutron star crust seems more likely.
According to previous studies (\citealt{HK09,HH12,BC18}), equation (\ref{e:sigmamax})
can be a reasonable estimate for polycrystals but 
numerical factor is uncertain, with typical values ranged around $0.01$ 
(it can be about twice smaller or larger). 
Therefore, our analysis of breaking conditions
in neutron star crust in the next sections should be
regarded as qualitative.

\section{Torsional oscillations of neutron star crust}
\label{s:torsion}

\subsection{Preliminaries}
\label{s:preliminary}

Let us apply the results of
Section \ref{s:deformed} to analyze torsional oscillations of neutron star crust.
We will consider the simplest model of a non-rotating and 
non-magnetic neutron star with crystalline crust and liquid core.
We will treat crustal matter as a
poly-crystal (isotropic solid) 
based on the Coulomb lattice of spherical atomic 
nuclei of one type at any  given density $\rho$. The density
can range from the density of solidified (e.g. iron) matter 
in the surface layers to about one half of the saturation density of nuclear
matter ($1.5 \times 10^{14}$ \gcc) at the crust-core interface
(e.g., \citealt{HPY2007}).

We will follow the theory of torsional oscillations developed 
in a very detailed paper by \citet{1983ST}. The authors have studied torsional  oscillations
taking into account associated space-time oscillations and emission of 
gravitational waves. Since these effects of General Relativity are tiny, we
restrict ourselves by the relativistic Cowling approximation, in which case space-time
is not perturbed and gravitational wave emission is neglected. This approximation
is well justified and greatly
simplifies theoretical consideration of the torsional oscillations.       

\subsection{General equations}
\label{s:equations}

Let us summarize the equations of \citet{1983ST} neglecting
space-time perturbations. The mertic inside and outside of a spherically 
symmetric non-perturbed star is taken in the standard form
\begin{equation}
 {\rm d s}^2=- {\rm e }^{2\Phi}\,{\rm d}t^2 + {\rm e}^{2\Lambda}\,{\rm d}r^2
  + r^2 ({\rm d}\theta^2 + \sin^2 \theta \, {\rm d}\phi^2),
\label{e:metric}
\end{equation}
where $t$ is Schwartzschild time (for a distant observer), $r$ is
a radial coordinate (circumferential radius), $\theta$ and $\phi$ 
are ordinary spherical angles, while $\Lambda$ and $\Phi$ are two metric
functions of $r$. At any $r$ one has
\begin{equation}
   \exp \Lambda (r)= \frac{1}{\sqrt{1-2Gm(r)/rc^2}},
\label{e:Lambda}   
\end{equation}
where $m(r)$ is the gravitational mass enclosed within a sphere of radius $r$ and 
$G$ is the gravitational constant.

Let $r=R$ be the radius of the star at which the pressure of the matter, $P(R)=0$.
Then $M=m(R)$ is the gravitational mass of the star. The functions $m(r)$, $P(r)$ and
$\Phi(r)$ within the star are obtained from the standard equations of hydrostatic
equilibrium [equations (5) in \citealt{1983ST}], supplemented by the equation of
state of stellar matter, which relates the pressure $P$ with the energy density
$\rho c^2$.  

Note the relation [equation (5h) in \citealt{1983ST}]
\begin{equation} 
 \Phi'+\Lambda'=4 \pi G c^{-1} (\rho+P/c^2)\exp(2\Lambda),
\label{e:phi+lambdaprime} 
\end{equation}
where prime denotes differentiation with respect to $r$.

Outside the star ($r>R$) one has
\begin{equation}
\exp \Phi= \exp(-\Lambda)= \sqrt{1 -r_{\rm g}/r},
\label{e:outsideNS}
\end{equation}
with $r_{\rm g}=2GM/c^2$ being the Schwartzschild radius.

\subsection{Torsional oscillations}
\label{s:torsional-oscillations}

Torsional oscillations  consist of shear 
motions of crystallized matter in the neutron star crust along 
spherical surfaces. In the adopted approximation,
they are not accompanied by oscillations
of density and pressure. The oscillations 
are weak and studied by linearizing the equations of
shear motion. Then one comes \citep{1983ST} to a set of
independent eigenmodes characterized by traditional 
spherical harmonics. Each mode can be
specified by multipolarity $\ell=2,3,\ldots$, 
azimuthal number ($-\ell \leq m_\ell \leq \ell$), as well as
by the number of radial nodes $n=0,1,\ldots$ 
The eigenfrequencies
$\omega=\omega_{\ell n}$ are naturally 
degenerate in $m_\ell$. In order to
find the oscillation spectrum, it is sufficient to set $m_\ell=0$. 
Then, among the three spacial
coordinates ($r$, $\theta$ and $\phi$) of a crystalline
matter element, only the angle $\phi$ varies.
A small proper spacial displacement of the matter element 
can be written
as
\begin{equation}
      {\rm d}l=r\,Y(r)\,\exp({\rm i}\omega t)b_\ell(\theta),
      \quad b_\ell(\theta)= \, 
      \frac{\partial}{\partial \theta} P_\ell(\cos \theta),
 \label{e:displace} 
\end{equation}
where $P_\ell(\cos \theta)$ is a Legendre polynomial. 
The dimensionless function 
$Y(r)$ is the radial part of the 
angular oscillation amplitude (to be calculated); it 
is real for our problem. A complex oscillating exponent
$\exp({\rm i}\omega t)$ has to be understood in
a standard way (as a real part). The function 
$b(\theta)$ describes the $\theta$-dependence 
of the oscillation amplitude. For instance, 
$b_2(\theta)=3\cos \theta\,\sin \theta$ and $b_3(\theta)=
1.5(5 \cos^2 \theta-1 )\,\sin \theta$. 
At any $\ell$ the vibrational motion vanishes along at
the `vibrational' axis $z$ [since one always has 
$b(\theta)\propto \sin \theta$].
If $\ell=2$, the vibrational motion is absent also at the
stellar equator; then crystal elements are shifting in
opposite directions in the upper and lower hemispheres of the star.
At $\ell=3$ vibrations disappear along conical
surfaces with $\cos^2 \theta=1/5$, which separate the oscillating 
crust into three zones. Recall that we are discussing 
the oscillation modes with $m_\ell=0$. Consideration
of the modes with $m_\ell \neq 0$ is similar and 
standard (using spherical harmonics). 

The equation for $Y(r)$ can be written as
\begin{align}
 Y''  +  \left( \frac{4}{r}+\Phi'-\Lambda'+\frac{\mu'}{\mu} \right) Y'
\nonumber \\
 + \left[ \frac{\rho+P/c^2}{\mu}\,\omega^2 {\rm e}^{-2\Phi}
  -\frac{(\ell+2)(\ell-1)}{r^2}\right]{\rm e}^{2\Lambda}Y=0.
\label{e:eqforY}    
\end{align}
It follows from equation (19b) of \citet{1983ST} and equation 
(\ref{e:phi+lambdaprime}) and presented in many 
publications (e.g., \citealt{2012Sotani}).
In order to determine the pulsation frequency, it has to be solved
with the boundary conditions $Y'(r_1)=0$ and $Y'(r_2)=0$ at both
(inner and outer) boundaries $r_{1,2}$ of the crystalline matter. 

Note that the quantity
\begin{equation}
     v_{\rm s}(r)=\sqrt{\frac{\mu(r)}{\rho+P(r)/c^2}},
\label{e:vsound}     
\end{equation}
in the square brackets of equation (\ref{e:eqforY})
is a local velocity of the radial shear wave as measured by a local observer.

\subsection{Formal relations}
\label{s:formal-relations}

Let us present 
some useful formal relations derived by \citet{1983ST}.

First of all, in the absence of dissipation, the vibrational energy 
of any mode, as measured by a distant observer, is
[equation (62a) in \citealt{1983ST}]
\begin{eqnarray}
 && E_{\rm vib}= \frac{\mathrm{\pi} \ell (\ell+1)}{2 \ell+1}\,
  \int_{r_1}^{r_2}{\rm d}r\, \left[ \omega^2 (\rho+P/c^2)\,{\rm e}^{\Lambda-\Phi}
  |Y|^2 \right.
\nonumber \\
&& \left. + \mu r^4 {\rm e}^{\Phi-\Lambda}|Y'|^2 
  + \mu (\ell+2)(\ell-1)r^2 {\rm e}^{\Phi+\Lambda}|Y|^2\right].
\label{e:energy}
\end{eqnarray} 
Then from equations (58a--58c) in the relativistic
Cowling approximation we obtain
\begin{eqnarray}
&& \omega^2= B/A,
\label{e:omegaBA} \\
&& A=\int_{r_1}^{r_2}{\rm d}r\, (\rho+P/c^2)\,{\rm e}^{\Lambda-\Phi}|Y|^2,
\label{e:omegaA} \\
&& B=\int_{r_1}^{r_2}{\rm d}r\, \left[\mu r^4 {\rm e}^{\Phi-\Lambda}|Y'|^2 \right.
\nonumber \\
&& \left.  + \mu (\ell+2)(\ell-1)r^2 {\rm e}^{\Phi+\Lambda}|Y|^2 \right].
\label{e:omegaB}
\end{eqnarray}
Clearly, $A$ and $B$ determine the vibrational energy $E_{\rm vib}$, and equation
(\ref{e:omegaBA}) can be used for formulating the variational principle for
torsional oscillations. 

Let us also present the expression for the viscous dissipation rate of 
the torsional vibration energy, $\dot{E}_{\rm vib}$, for a distant observer (where
dot means differentiation over the Schwartzschild time). It is easily
derived using the formalism of \citet{1983ST},
\begin{equation}
 \dot{E}_{\rm vib}= \frac{\mathrm{\pi} \ell (\ell+1)}{2 \ell+1}\,\omega^2
\int_{r_1}^{r_2}{\rm d}r\, 
\eta r^2 {\rm e}^{-\Phi-\Lambda}|Y'|^2,
\label{e:Edot}
\end{equation}
where $\eta$ is the shear viscosity in a neutron star
crust \citep{2005Chugunov,2008Shternin}. 
We naturally assume that the dissipation time is much 
slower than vibration period.

\section{Properties of torsional oscillations}
\label{s:properties}

\subsection{Neutron star model}
\label{s:NSmodel}

\begin{figure}
\includegraphics[width=0.45\textwidth]{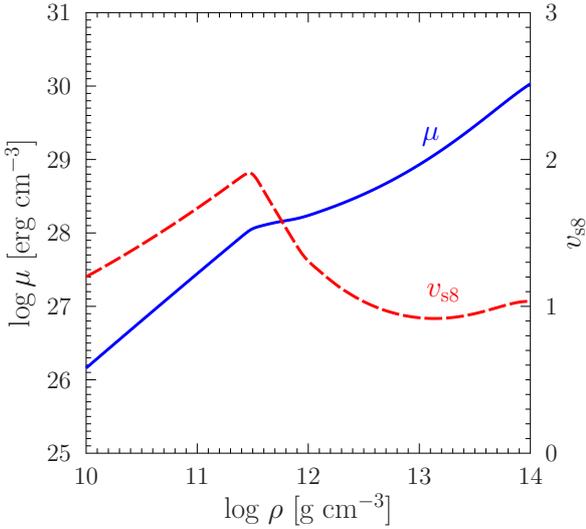}%
\hspace{5mm}
\caption{The shear modulus $\mu$ (the left-hand vertical axis) and the 
local shear velocity $v_{\rm s}$ (in units of $10^8$ cm s$^{-1}$, the
right-hand vertical axis) as a function of density in 
the neutron star crust. 
}
\label{f:murho}
\end{figure}

For an example, we have chosen a typical model of a neutron star with
$M=1.4\, \msun$ and nucleon core. The equation of state  
is based on the results 
obtained by \citet{1998APR}. It was suggested by \citet{2005Gusakov} 
(and called APR III) using 
a numerical parameterization of \citet{1999HHJ,2000HHJ}. The
stellar radius is $R=12.27$ km. The neutron drip density 
$\rho_{\rm drip}=4.3 \times 10^{11}$ \gcc\ is reached
at $r_{\rm drip}=11.82$ km and $m_{\rm drip}=1.3998\, \msun$. 
The crust-core interface is placed at $\rho_{\rm cc}=1.45 \times
10^{14}$ \gcc\ (with $r_{\rm cc}=10.09$ km and $m_{\rm cc}=1.364\, \msun$).
The shear modulus $\mu$ in the cold catalyzed crust is calculated from equation
(\ref{e:mu}) using the smooth composition model of spherical
nuclei \citep{HPY2007}. Fig.\ \ref{f:murho} shows the dependence
of $\mu$ on $\rho$ (the left-hand vertical axis), as well as 
the dependence of the local shear-wave velocity $v_{\rm s}$ on $\rho$ 
[equation (\ref{e:vsound}), the right-hand axis].
Note that although the smooth composition model averages 
density variations of $\mu$ (produced by variations of nuclear
parameters with depth), it formally gives a small jump of
$\mu$ at the neutron drip point. On the other hand, the problem
of accurate calculation of elastic moduli in the deep crust 
is far from being solved \citep{HPY2007} and the exact $\mu(\rho)$ 
dependence is actually not known. For that reason, we have smoothed out artificially
the $\mu(\rho)$ jump at the neutron drip point; we will comment on
theoretical uncertainties of $\mu(\rho)$ below. In any case 
Fig.\ \ref{f:murho} demonstrates different behavior of $\mu$ and $v_{\rm s}$ 
in the outer and inner crust (separated by the 
density $\rho_{\rm drip}$).

\subsection{Eigenfrequencies and eigenmodes}
\label{s:eigenproblem}

Let us analyze torsional oscillation modes $(\ell,n)$ described 
by equation (\ref{e:eqforY}). The cyclic oscillation frequencies will be
denoted as 
$\nu_{\ell n}=\omega_{\ell n}/(2 {\rm \pi})$. 
Equation (\ref{e:eqforY}) is easily solved numerically. A solution 
can be specified by $Y(r_2)=Y_0$ at the outer boundary
of the crystalline shell. Any solution gives an eigenfrequency
$\nu$ as well as radial functions $Y(r)$ and $Y'(r)$. The first
function determines horizontal shifts of crystalline matter
elements (\ref{e:displace}), while the second function 
specifies the only one non-trivial 
strain tensor element of the problem, $u_{r \phi}$. In a local 
reference frame we have
\begin{equation}
   u_{r \phi}= \frac{r Y'(r)}{2 \exp \Lambda(r)}
   \,b(\theta).
\label{e:strain}     
\end{equation}

It is convenient to
characterize $u_{r \phi}$ by the radial strain function 
defined as
\begin{equation}
   \xi(r) \equiv \frac{r Y'(r)}{2 Y_0 \exp \Lambda(r)}.
\label{e:xi}   
\end{equation}
It is independent of the normalization constant $Y_0$ and
shows how strained the crystal is in a local matter
element. 

Equation (\ref{e:strain}) allows us to write down the
only one non-trivial elastic stress tensor component of 
oscillating crystal,
\begin{equation}
    \sigma_{r\phi}=
     \frac{\mu r Y'(r)}{ \exp \Lambda (r)}\,b(\theta),
\label{e:sigma}     
\end{equation}
and formulate the local crystal breaking condition.
The breaking occurs if $|\sigma_{r \phi}|$ exceeds the 
breaking stress $\sigma_{\rm max}$. Comparing equations
(\ref{e:sigmamax}) and (\ref{e:mu}) we have
$\sigma_{\rm max}=\alpha_*\, \mu$,
where $\alpha_* \approx 0.02/0.1194\approx 0.17$ is a constant 
provided by the theory of deformed Coulomb crystals (Section \ref{s:deformed}). 
The breaking 
occurs in a point $r_{*}$ at an angle $\theta_{*}$ if
the pulsation amplitude $Y_0$ is sufficiently large,
\begin{equation}
  |Y_{0*}| \geq \frac{\alpha_* }{2b(\theta_*)|\xi(r_{*})|}.
 \label{e:break}
\end{equation}
Here $r_{*}$ refers to the maximum of $|\xi(r)|$, and
$\theta_{*}$ to the maximum $b_{\rm max}$ of $|b(\theta)|$.
For $\ell=2$ we have $\theta_{*}={\rm \pi}/4$ and
$b_{\rm max}=3/2$, while for $\ell=3$ we obtain
$\sin^2 \theta_{*}=4/15$ and 
$b_{\rm max}=8/ \sqrt{15}\approx 2.066$.

The properties of fundamental torsional modes (without radial nodes, $n=0$)
and ordinary modes (with radial nodes, $n>0$) are drastically 
different (e.g., \citealt{1983ST}) and will be described separately.
In Table \ref{tab:param} we present 
numerically calculated frequencies $\nu$ and 
vibration energies $E_{\rm vib}$ for two fundamental 
modes ($\ell=2$
and 3; $n=0$) and four ordinary ones ($\ell=2$ and $3$; 
$n=1$ and 2). The three last columns in Table \ref{tab:param} are
concerned with crystal breaking as detailed in the next sections.

\subsection{Fundamental torsional oscillations}
\label{s:fundamental}

These oscillations possess remarkable properties
associated with the fact that the elastic shear modulus
in a neutron star crust is much smaller than the
compressional modulus (the latter determines hydrostatic
structure of the crust). In the absence of radial
nodes, the vibrating crystal remains strongly 
understrained (non-deformed), with $|\xi(r)| \ll 1$ in the entire crystalline
shell. This means that the crystal is almost fully relaxed,
with
\begin{equation}
  Y(r)\approx Y_0
\label{e:Yconst}
\end{equation}
anywhere at $r_1 \leq r \leq r_2$.

\renewcommand{\arraystretch}{1.2}
\begin{table}
\caption{Some torsional oscillation parameters for  a $1.4\,\msun$ neutron star
($Y_0$ is expressed in radians); see the text for details}
\label{tab:param}
\begin{tabular}{c c c c c c}
		\hline 
$\ell$, $n$	& $\nu$ [Hz] & $E_{\rm vib}$ [erg] & $|\xi_*|^{a)}$ & $Y_{0*}^{b)}$ & $E_{\rm vib}^{*c)}$ [erg]  \\
		\hline
2, 0   & 22.77 &  $1.64 \times 10^{48}\,Y_0^2$ & 0.01 & $\ll 1$ &$\ll 2 \times 10^{48}$\\
3, 0   & 36.01 &  $5.88 \times 10^{48}\,Y_0^2$ & 0.03 & $\ll 1$ &$\ll 6 \times 10^{48}$\\
2, 1   & 631.1  & $5.30 \times 10^{49}\,Y_0^2$ & 9.7 & 0.006 & $2 \times 10^{45}$\\
3, 1   & 631.6 &  $7.60 \times 10^{49}\,Y_0^2$ &9.7 & 0.004 & $1.3 \times 10^{45}$\\
2, 2   & 1031.3  & $1.66 \times 10^{49}\,Y_0^2$ & 10 & 0.006 & $5 \times 10^{44}$\\
3, 2   & 1031.6 &  $2.38 \times 10^{49}\,Y_0^2$ & 10 & 0.004 & $4 \times 10^{44}$\\
\hline
\end{tabular}	
 \\
	$^{a)}$ Breaking value of radial strain function \\
	$^{b)}$ Breaking or limiting angular vibration amplitude \\
	$^{c)}$ Breaking or limiting vibrational energy  
\end{table}
\renewcommand{\arraystretch}{1.0} 

\begin{figure}
	\includegraphics[width=0.45\textwidth]{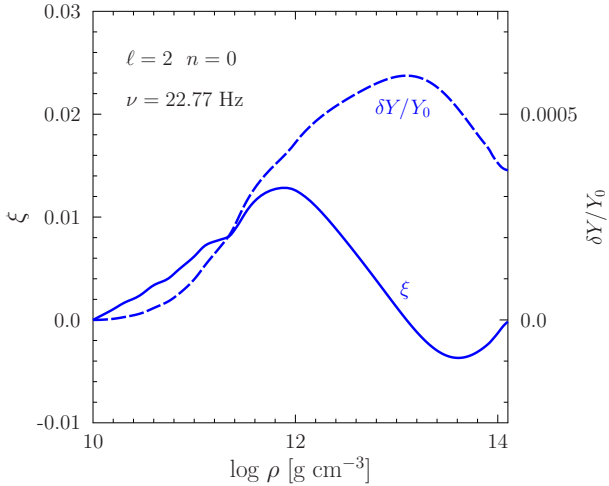}%
	\hspace{5mm}
	\caption{Radial eigenfunctions versus density $\rho$ in the neutron star 
	crust for the mode with $\ell=2$ and $n=0$ at $\nu=22.77$ Hz. The solid line is the 
	radial strain function (\ref{e:xi}) that is plotted along the
	left-hand vertical axis. The dashed line demonstrates 
	very weak relative deviations $\delta Y/Y_0$ of $Y(\rho)$ from $Y_0$ [plotted
	along the right-hand  vertical axis, with $\delta Y=Y(\rho)-Y_0$]. 
	}
	\label{f:dif20}
\end{figure}

For instance, Fig.\ \ref{f:dif20} shows the  radial dependence of 
$Y(r)$ in the crust of the $1.4\,\msun$ 
star (Section \ref{s:NSmodel}) for the simplest
torsional oscillation mode with $\ell=2$ and $n=0$.
Instead of the radial variable $r$ we use the density
variable $\rho$. The outer
crystalline boundary is placed at $\rho=10^{10}$ \gcc\ (about 165 m under
the surface) and the highest density corresponds to the
crust-core interface. We will show that shifting
the outer boundary to the surface or inside of the star does not affect the results.
The solid line shows (the left-hand
vertical axis) the radial strain function
$\xi$ defined by equation (\ref{e:xi}). We see that the
values of $\xi$ are indeed small, $|\xi(\rho)|\lesssim 0.01$, so that
the crystal stays nearly non-deformed. As a result, the
approximation (\ref{e:Yconst}) is almost perfect.
Relative deviations of $Y(\rho)$ from $Y_0$ 
are shown on the same Fig.\ \ref{f:dif20} by the dashed line
along the right-hand axis. They do not exceed 0.0006. 

Note some wiggles of the $\xi(\rho)$ curve near the
neutron drip density (Fig.\ \ref{f:dif20}). They result from 
our artificial smoothing of the $\mu(\rho)$
dependence described in Section \ref{s:NSmodel}. Making 
more accurate smoothing, we could remove the wiggles. 
We have left them as a reminder that the reality 
is usually more complicated than artificially smoothed 
theoretical curves. The problem of constructing a realistic
$\mu(\rho)$ dependence is complicated. On general grounds,
one can expect that this dependence is not smooth, especially
at those densities, where one preferable nucleus type is 
replaced by another (e.g., \citealt{HPY2007}). This would lead to wiggling of $\xi(\rho)$
[and, to a less a extent, of $Y(\rho)$] curves, which may affect
torsional oscillations (especially with high $n$ and $\ell$).  

Having $\xi(r)$ from Fig.\ \ref{f:dif20}, we can check the 
crystal breaking condition (\ref{e:break}) for the 
($\ell=2,~n=0$) mode. The breaking point
would be at $\theta_{*}={\rm \pi}/4$ and $\rho_{*}\approx
10^{12}$ \gcc. However, since $|\xi(\rho)|\ll 1$, the
breaking would require large oscillation amplitudes, $Y_0 \gg 1$,
which are beyond the linear oscillation theory.
Accordingly, the fundamental  torsional 
oscillations do not break in the linear regime (at $|Y|\ll 1$).  

\begin{figure}
	\includegraphics[width=0.45\textwidth]{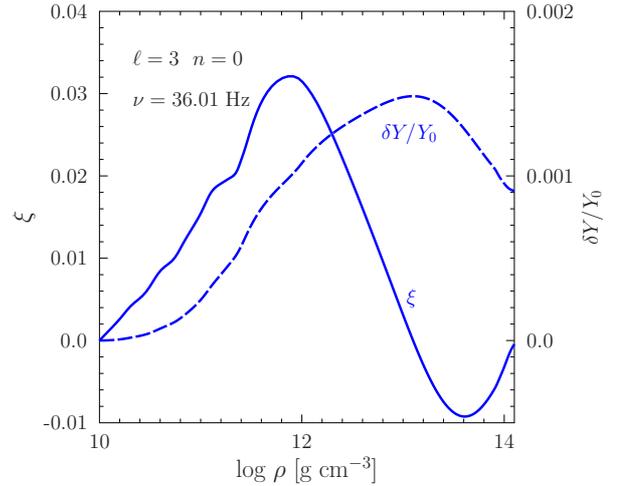}%
	\hspace{5mm}
	\caption{The same as in Fig.\ \ref{f:dif20} 
		but for the mode with $\ell=3$ and $n=0$ at 
		 $\nu=36.01$ Hz. 
	}
	\label{f:dif30}
\end{figure}

This conclusion holds for all fundamental oscillations
which keep the  crystal almost relaxed.
For instance, Fig.\ \ref{f:dif30} shows the same
functions $\xi(\rho)$ and $\delta Y(\rho)/Y_0$ 
for the next fundamental mode
with $\ell=3$ and $n=0$. The results are  similar
although the approximation of almost full relaxation becomes
worse with increasing $\ell$ (the values of $|\xi|$
are getting higher). 

The same approximation 
greatly simplifies calculations of the
eigenfrequencies. Indeed, in this case we can use equation 
(\ref{e:omegaBA}), neglect the terms containing $Y'$ in 
equations (\ref{e:omegaB}) and (\ref{e:omegaA}) and set
$Y=Y_0$ there. Then we obtain
\begin{eqnarray}
&& \omega_{\ell 0}^2
=\frac{1}{4}\omega_{20}^2 \,(\ell+2)(\ell-1),\quad
\omega_{20}^2= \frac{8 E_\mu}{3 I_{\rm cr}},
\label{e:omegaBA0} \\
&& I_\text{cr}=\frac{8 {\rm \pi}}{3}\int_{r_1}^{r_2}{\rm d}r\,r^4 (\rho+P/c^2)\exp(\Lambda-\Phi),
\label{e:omegaA0} \\
&& E_\mu = 4 {\rm \pi} \int_{r_1}^{r_2}{\rm d}r\, r^2 \mu \exp(\Phi+\Lambda).
\label{e:omegaB0}
\end{eqnarray}
This is an explicit analytic expression for the
pulsation frequencies $\omega_{\ell 0}$ in terms of two simple one-dimensional
integrals, $I_{\rm cr}$ and $E_\mu$, which are easily  computed once a neutron star model and
crustal microphysics are known. If we set $\exp \Phi(r)=1$ under the integrals (as if we
 ignore gravitational time delay of signals for a distant observer), we would
immediately identify
$I_{\rm cr}$ as the moment of inertia of the crust and
$E_\mu$ as the shear modulus integrated over the crust (which is close to 
the total electrostatic Coulomb energy of the crust). The factor $\exp \Phi(r)$ 
under the integrals is needed to express $\omega$ for a distant observer.  
To the best of our knowledge, this explicit representation of
torsional fundamental pulsation frequencies
has not been known in the literature. 

In our case, $E_\mu=6.85 \times 10^{47}$ erg
and $I_{\rm cr}=8.91 \times 10^{43}$ g~cm$^2$. We have checked that 
equation (\ref{e:omegaBA0}), indeed, accurately reproduces
the frequencies $\nu_{\ell 0}$ of fundamental modes
derived by solving the basic differential
equation (\ref{e:eqforY}).

According to equation (\ref{e:omegaBA0}), the spectrum of 
fundamental modes is expressed through the lowest 
frequency $\omega_{20}$ and multipolarity $\ell$ and behaves
as $\omega_{\ell 0} \propto \sqrt{(\ell-1) (\ell+2)}$, which gives, for instance, 
$\omega_{30}/\omega_{20}=\sqrt{2.5}=1.581$. At large
$\ell$ the neighboring frequencies become equidistant.
However, some publications predict, from 
qualitative considerations, another scaling 
$\omega_{\ell 0} \propto \sqrt{\ell (\ell+1)}$ 
[e.g. equation (3a) in \citealt{1983ST}, or equation
(3) in \citealt{2016Sotani}]; it gives 
$\omega_{3 0}/\omega_{20} = \sqrt{2}=1.414$. We see that
the exact form is somewhat different from the approximate
one. 

We should stress that the correctness of the scaling 
$\omega_{\ell 0} \propto \sqrt{(\ell-1) (\ell+2)}$ has
been discussed in the literature, although it has not 
been written 
in an explicit form (\ref{e:omegaBA0}). In particular,
\citet{2007Samuel} present very convincing 
arguments (based on many numerical
calculations) that  the ratio $\omega_{30}/\omega_{20}=\sqrt{2.5}$
is true. \citet{2016Gabler} in their table 2 
present the frequencies $\omega_{\ell 0}$ with $\ell$ from 2 to 6,
which agree with the correct scaling (except for $\ell=6$, where the frequency contains the typo and 
should be 83.8 Hz). However,
\citet{2016Sotani} prefer to use the approximate scaling
to fit the fundamental oscillation frequencies.

The approximation of almost relaxed crystal allows
us to use equation (\ref{e:energy}) 
and obtain the energy of any fundamental mode in analytic form,
\begin{equation}
E_{\rm vib}^{\ell 0}=
\frac{ (\ell-1)\ell (\ell+1)(\ell+2)}{2 (2 \ell +1)}\,E_\mu Y_0^2.
\label{e:energy0}
\end{equation}
This formula seems original as well.
The vibration energy is determined by the Coulomb energy 
of the crust and by the angular 
vibration amplitude
$Y_0$ (expressed in radians).
For instance, for the mode with $\ell=2$ and $n=0$ we have
$E_{\rm vib}^{2 0}\approx 1.64 \times 10^{48}\,Y_0^2$ erg, meaning that 
the energy can be quite substantial. Since the applicability of 
this expression is limited by the linear vibration regime ($Y_0 \ll 1$),
we conclude that the expression is valid as long as
$E_{\rm vib}^{2 0} \ll 2 \times 10^{48}$ erg.

\begin{figure}
	\includegraphics[width=0.45\textwidth]{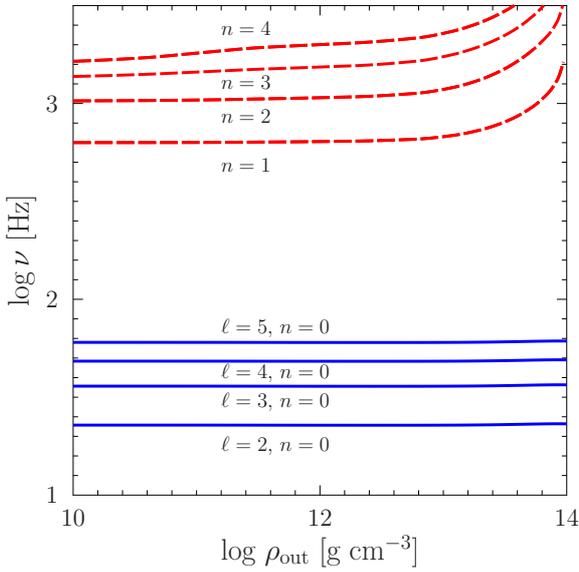}%
	\hspace{5mm}
	\caption{Spectrum of torsional oscillations localized in 
		an artificial crust, where the matter is
		crystallized only within the shell located
		starting from some outer density $\rho_{\rm out}$
		to the crust-core interface, as a function of $\rho_{\rm out}$.		
		Four lower solid lines refer to fundamental modes with
		 $\ell=2,3,4,5$ (from bottom to top).
		The dashed lines refer to ordinary modes with
		$n=1,2,3,4$. Each dashed line encloses a number of modes
		with $\ell=2,3,4$, whose frequencies are non-distinguishable
		in the logarithmic scale.  
	}
	\label{f:omegarho}
\end{figure}

Another remarkable property of torsional 
vibrations is shown in Fig.\ \ref{f:omegarho}. 
This figure exhibits the spectrum of 
oscillations with $\ell=2$, 3 and 4 and
$n=0,\ldots 4$. The frequencies 
are computed from the differential equation (\ref{e:eqforY}) 
by artificially assuming
that the crystallized shell occupies some fraction
of the crust, starting from an arbitrary
density $\rho_{\rm out}$ and ending at the
crust-core interface (as if the layer 
at $\rho< \rho_{\rm out}$ is 
melted). The frequencies are plotted
in logarithmic scale versus $\log \rho_{\rm out}$.

Four lowest solid lines refer to the family of fundamental
modes ($\ell$=2, 3, 4, 5 with $n=0$) which are  approximately
equidistant in the linear scale. These frequencies
are well described by equation (\ref{e:omegaBA0}). Their most interesting
feature is that they are almost insensitive to $\rho_{\rm out}$:
an artificial melting of any outer layer of the crust
has almost no effect on the spectrum of fundamental
modes. This feature seems to be closely related to the almost relaxed
state of crystals in fundamental oscillations. 

The dashed lines on Fig.\ \ref{f:omegarho} correspond to ordinary torsion 
modes with $n=1,\ldots 4$ (from bottom to top). Each dashed curve
shows actually a family of different modes with $\ell=$2, 3 and 4, but the 
`fine splitting' of the curves with different $\ell$ is so small 
that it is invisible in the logarithmic format. These modes will be
analyzed below.

\subsection{Ordinary torsional oscillations}
\label{s:withnodes}

\begin{figure}
	\includegraphics[width=0.45\textwidth]{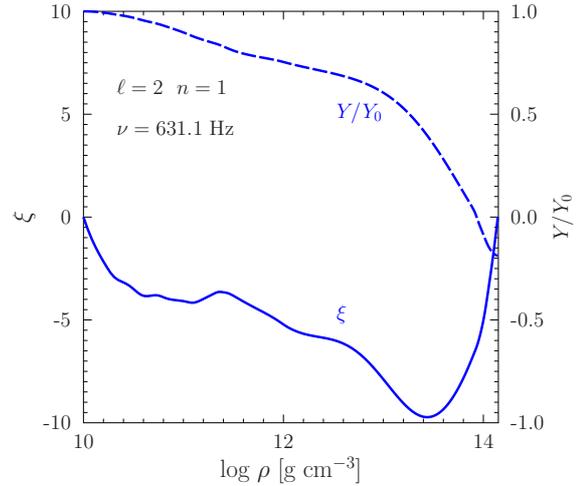}%
	\hspace{5mm}
	\caption{Radial eigenfunctions versus density $\rho$ in the neutron star 
		crust for the ($\ell=2,~n=1$) mode with 
		the frequency $\nu=631.06$ Hz. The solid line is the 
		radial strain function (\ref{e:xi}) that is plotted along the
		left-hand vertical axis. The dashed line demonstrates 
		variations of $Y(\rho)/Y_0$ 
		(the right-hand  vertical axis). 
	}
	\label{f:dif21}
\end{figure}

\begin{figure}
	\includegraphics[width=0.45\textwidth]{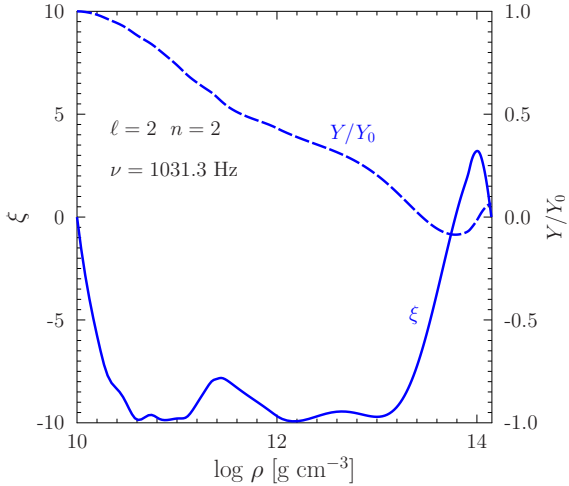}%
	\hspace{5mm}
	\caption{Same as in Fig.\ \ref{f:dif21} but
		for the ($\ell=2,~n=2$) mode with 
		$\nu=1031.3$~Hz.  
	}
	\label{f:dif22}
\end{figure}

Now let us turn to torsional oscillations with radial nodes ($n>0$). 
They are less remarkable.  
The main difference from the fundamental modes is that now 
crystalline matter is rather strained and stressed. 
The analytic expressions for the pulsation frequencies (\ref{e:omegaBA0})
and energies (\ref{e:energy0}) become inapplicable; one should calculate
these quantities from more complicated equations (\ref{e:eqforY})
and (\ref{e:energy}).

For example,
Fig.\ \ref{f:dif21} shows the radial strain function $\xi(\rho)$ 
(the left-hand vertical
axis) and the function $Y(\rho)/Y_0$ (the right-hand vertical axis) 
for the simplest ($\ell=2,~n=1$) mode. One can observe
single radial node of $Y(\rho)$ in the inner crust
at $\rho= 8.6 \times 10^{13}$ \gcc. A comparison with
Fig.\ \ref{f:dif20} (for $\ell=2,~n=0$) shows that
now the radial strain
function $|\xi(\rho)|$ is much larger and
reaches the values $|\xi| \sim 10$.   
The vibrating crust becomes dynamically deformed, 
which increases pulsation frequencies 
($\nu_{21}=631.06$ Hz versus $\nu_{20}=22.77$ Hz; see Table \ref{tab:param}).
For the same angular pulsation amplitude $Y_0$, 
the pulsation energy $E_{\rm vib}^{21}$
is about 30 times higher than $E_{\rm vib}^{20}$. 

Since the ($\ell=2,~n=1$) mode
is strained, it can break in the linear pulsation regime.
The breaking occurs (Fig.\ \ref{f:dif21}) in the inner crust 
at $\rho_* \approx 2.8 \times 10^{13}$ \gcc, where
$|\xi(\rho)|$ reaches maximum, $|\xi_*|\approx 9.7$, provided  
the pulsation amplitude  reaches the breaking value 
$|Y_{0*}|\approx 0.0058$ given by equation (\ref{e:break}). 
This crystal breaking limits
the maximum vibrational energy to $E_{\rm vib}^* \approx 6 \times 10^{47}$ erg
 in the linear pulsation regime.

Recall that the pulsation frequencies $\nu_{\ell k}$ of modes with given $n>0$ slightly 
increase with $\ell$ 
(Table \ref{tab:param}, Fig.\ \ref{f:omegarho}), demonstrating the `fine
splitting' structure (which was pointed out long ago; e.g.,
\citealt{1980Hansen}). The deformation profiles $Y(\rho)$ and $\xi(\rho)$ for such 
modes, as well as breaking densities $\rho_*$, are also slightly dependent of $\ell$. Our estimates
of breaking vibration amplitudes $|Y_{0*}|$ and vibrational energies
$E_{\rm vib}^*$ for some modes are presented in Table~\ref{tab:param}.

Fig.\ \ref{f:dif22} shows $Y(\rho)$ and $\xi(\rho)$ for the mode with $\ell=2$ and $n=2$. 
This mode is more complicated, with the two nodes of $Y(\rho)$
at $\rho\approx 2.8 \times 10^{13}$ and $1.03 \times 10^{14}$ \gcc\
near the crust-core interface. The crystal is more  
stressed there, which rises the oscillation frequency to $\nu=1.0313$ kHz.
Such oscillations of $Y(\rho)$ are becoming sensitive to small uncertainties
(wiggles) of the shear modulus $\mu(\rho)$ (discussed above); their
breaking can occur in the outer crust. Therefore, 
theoretical construction of high-frequency torsional oscillations can
be dependent on largely unknown details of the shear modulus behavior.

Now we return to the spectra of oscillation frequencies in a
theoretical experiment of artificial `melting' the outer layer of the crust 
(from the surface to
$\rho_{\rm out}$; Fig.\ \ref{f:omegarho}). In contrast to the fundamental
modes, the vibration frequencies of ordinary modes 
(dashed curves) may considerably depend on $\rho_{\rm out}$. 
Nevertheless, this dependence is 
most pronounced at sufficiently high $\rho_{\rm out}$, when the `melting' reaches
the inner crust. Shallower melting has almost no effect on $\nu_{\ell n}$.
It looks as if the outer crust does not affect the
torsional oscillation frequencies at all, so that theorists can
remove it  while calculating $\nu_{\ell n}$. Its only duty is to
allow the matter to vibrate near the surface for making the vibrations observable.

This effect has a simple explanation. The oscillations
are definitely controlled in the inner crust, where crystalline matter can be strongly 
stressed. In contrast,  oscillations of the outer 
crystalline layers remain rather relaxed and do not 
affect $\nu_{\ell n}$. Accordingly, the oscillation frequencies 
are almost insensitive to the physics of the outer crust (cold-catalyzed or accreted matter, 
mono-crystal or isotropic solid, one ion component or mixture of 
different ions, exact position of outermost melted 
layers and type of outer boundary condition).  

\section{Discussion and conclusions}    
\label{s:conclude}

We have studied deformed crystalline 
matter in neutron star crust. Firstly, in Section \ref{s:deformed}
we have performed semi-analytic
calculations of the breaking stress of Coulomb mono-crystals of atomic nuclei
under shear deformations of two types, (\ref{e:trans1}) and (\ref{e:trans2}), 
taking into account the electron plasma screening of Coulomb 
forces between the nuclei. 
We have fitted the results by the analytic expression (\ref{e:fit}). We have also 
analysed the breaking stress of Coulomb crystals in the model of isotropic
solid, which is needed for applications.

Secondly, we have presented some new results on torsional oscillations 
of spherically symmetric and non-magnetic neutron star crust.
In particular, we have derived exact analytic expressions (\ref{e:omegaBA0})
and (\ref{e:energy0}) for frequencies and energies of fundamental torsional
oscillations, analyzed the same quantities for ordinary torsional
oscillations, and pointed out a specific independence of the torsional
vibrations of physical conditions in the outer crust. 
In addition, we have formulated the conditions for breaking the
torsional oscillations. The torsional pulsation spectra have been extensively
studied before, but pulsation energies, distribution of deformations
and stresses over oscillating crust, and the breaking conditions have
been given little attention.  

The spectra of torsional oscillations in 
spherical non-magnetic stars have been 
analyzed in many publications  
(e.g., \citealt{2009Andersson,2012Sotani,2013aSotani,2013Sotani,2016Sotani},
and references therein). The authors have used different 
modern equations of state of neutron star matter and wide ranges
of neutron star masses with the aim to explore the sensitivity 
of oscillation spectra on the equation of state and 
nuclear physics of the matter near the crust-core interface
(the density dependence of the symmetry energy, possible presence
of nuclear pasta phases, superfluidity of neutrons). We have intentionally
chosen one equation of state and one neutron star mass because the 
effects of different equations of states and masses have already
been explored. 

So far, the theory of torsional oscillations has been employed to interpret 
spectra of QPOs observed in X-ray tails of flares of three SGRs. SGRs belong to
a class of magnetars;
their flaring activity is regulated by very strong magnetic fields,
$B \sim 10^{15}$ G 
(see, e.g., \citealt{2015Mereghetti,2017KasB}, for a recent review). These
QPOs were detected in the hyperflare 
of SGR 1806--20 (in 2004), in the giant flare SGR 1900+14 (1998), and 
in sequences of numerous less energetic recurrent bursts of
SGR 1806--20 (1996) and SGR J1550--5418 (2009) 
(e.g., \citealt{2005Israel,2006Watts,2011Hambaryan,2014Huppen,2014Huppenkothen},
and references therein). 
For example, the frequencies of QPOs discovered in the afterglow of the
hyperflare of SGR 1806-20 are 
18, 26, 30, 92, 150, 625, 1840 Hz and (with somewhat less confidence) 17, 21, 36,
59, 116 Hz.  The QPO frequencies detected in other events range from 28 to
260 Hz.

All these QPOs can (in principle) be explained by
torsional oscillations of spherical non-magnetic neutron stars. 
However, the current state of the
art in neutron star physics strongly suggests to include  
important effects of high magnetic fields, superfluidity of nucleons 
and other effects of nuclear interactions (like entrainment) in the
stellar core.
There have been many investigations of global low-frequency 
oscillations of magnetized superfluid neutron stars 
(e.g., \citealt{2006Levin,2007Levin,2006Glampedakis,2007Sotani,2009CD,2009Colaiuda,2011Colaiuda,2012Colaiuda,2011vanHoven,2012vanHoven,2011Gabler,
2012Gabler,2013Gabler,2013Gabler1,2016Gabler,2018Gabler,
2014Passamon,2016Link}). A comprehensive discussion of
the current state of the affair can be found in \citet{2016Gabler,2018Gabler}.
The two major ingredients of the theory are 
torsional crustal oscillations (affected
by crustal magnetic fields) and continuum of Alfv\'en shear waves 
in the core. Interaction of these waves leads to the appearance
of discrete spectrum of magneto-elastic oscillations. The spectrum 
essentially depends on the magnetic field strength and geometry
and on superfluid properties of neutron star matter.
These oscillations can also be coupled with MHD oscillations in 
magnetospheres of magnetars (e.g., \citealt{2016Link}).

The theory enables one to interpret QPOs in SGRs, 
but possible interpretations seem still not 
unambiguous because of vast diversity of theoretical models and 
associated degeneracy in the parameter space. 
Some theoretical problems have not been considered yet, for instance, possible effects
of hyperons in neutron star cores and many sophisticated  relativistic MHD effects
(e.g. \citealt{2020Dommes}). 
This motivates further studies of magneto-elastic oscillations.

Our results can be useful for describing these oscillations 
in the crust. Note that we have not touched the problem of oscillation 
excitation and damping, as well as the 
transformation of the oscillations into observable QPOs. Torsional
oscillations in idealized spherical non-magnetic 
stars could live for a long time. Their decay times due to gravitational 
radiation were estimated \citep{1983ST} to be about $10^4$ years, and 
damping due to shear viscosity  calculated from equation (\ref{e:Edot})
would be even much longer. In contrast, QPOs in SGR flares 
loose coherence in a fraction of second, which suggests QPOs cannot
be just ordinary torsional crustal oscillations. 
There are many channels of fast damping of oscillations
in magnetized stars discussed in the references cited above (e.g., \citealt{2011Gabler}).
Also, there could be different mechanisms for excitation of the magneto-elastic
oscillations in flaring SGRs (e.g. \citealt{2016BL}). 

Let us stress that magneto-elastic oscillations in
SGRs can be very different from classical torsional oscillations
in non-magnetic crust. Superflares of SGRs are accompanied
by enormous energy release (e.g. \citealt{2016BL}), making the crust hot and melting
some regions, especially outer layers, where the melting
temperature is sufficiently low. According to our results (Fig.\ \ref{f:omegarho}), 
such a melting does not affect the oscillation frequencies,
which supports the idea that QPOs are connected with magneto-elastic
oscillations. However, an additional study is required to
show that our results remain valid under SGR conditions.

Our results can also be useful to simulate other processes involving 
deformed crystals in neutron stars like evolution of
the crustal magnetic fields \citep[e.g.,][]{BL14,L16,LL16}
and pulsar glitches \citep[e.g.,][]{PFH14}.   

\section*{Acknowledgments}
We are grateful to D. Baiko for providing us the
data on deformed Coulomb crystals and to the
anonymous referee for pointing out that 
our results may support the connection of QPOs in
flaring SRGs with magneto-elastic
oscillations.
The work was supported by the Russian
Science Foundation (grant 19-12-00133). 

\section*{Data availability}
The data underlying this article will be shared on
reasonable request to the authors.

\bibliographystyle{mnras}

\bibliography{YakBibList}

\end{document}